\documentclass[useAMS,usenatbib]{mn2e}

\usepackage{times}
\usepackage{graphicx}
\usepackage{rotating}

\title[Proper motions of field L and T dwarfs]{Proper motions of field L and T dwarfs -II}
\author[S. L. Casewell et al.]{S. L. Casewell$^{1}$\thanks{E-mail:
slc25@star.le.ac.uk}, R. F. Jameson$^{1}$ and M. R. Burleigh$^{1}$ \\
$^{1}$Department of Physics and Astronomy, Univeristy of Leicester, Univeristy Road, Leicester, LE1 7RH, UK\\
}
\begin{document}

\date{}

\pagerange{\pageref{firstpage}--\pageref{lastpage}} \pubyear{2007}

\maketitle

\label{firstpage}

\begin{abstract}
By using images taken with WFCAM on UKIRT and SofI on the NTT and combining them with 2MASS we have measured proper motions for 126 L and T dwarfs in the dwarf archive. Two of these L dwarfs appear to have M dwarf common proper motion companions, and 2 also appear to be high velocity dwarfs, indicating possible membership of the thick disc.
We have also compared the motion of these 126 objects to that of numerous moving groups, and have identified new members of the Hyades, Ursa Major and Pleiades moving groups.
These new objects, as well as those identified in \citet{jameson08} have allowed us to refine the L dwarf sequence for Ursa Major that was defined by \citet{jameson08b}.
\end{abstract}


\section{Introduction}
Brown dwarfs may be thought of as failed stars. These low mass ($\leq$70 M$_{\rm Jup}$ \citealt{burrows01}), cool objects are the lowest mass 
objects that 
the star formation process can produce. The majority of the brown dwarfs that have been discovered to date are field objects found
using surveys such as the Two Micron All Sky Survey (2MASS; \citealt{skrutskie06}, see \citealt{leggett02} for examples), the DEep Near-Infrared 
Sky survey (DENIS; \citealt{denis05}, see \citealt{delfosse99} for examples), the Sloan Digital Sky Survey (SDSS;\citealt{york00} see \citealt{hawley02} 
for examples) and the UKIRT Deep Infrared Sky Survey (UKIDSS; \citealt{lawrence06}, see \citealt{kendall07, lodieu07b} for examples). 
However, to study brown dwarfs in depth, a knowledge of their age is essential, which means we must study brown dwarfs in open star clusters 
or moving groups.

Once a brown dwarf has been proved to belong to an open star cluster, or a moving group, then its age is known, 
allowing meaningful comparisons to evolutionary models to be made. The most recent example of this is the study done by 
\citet{bannister07} who used existing proper motions and parallax measurements to show that a selection of field dwarfs in fact belong 
to the Ursa Major and Hyades moving groups. The importance of this study, is that these are the first brown dwarfs to be associated 
with an older cluster or group (age $>$200 Myr). Older clusters such as the Hyades are expected to contain very few or no brown dwarfs or low mass
 members, due to the dynamical evolution of the cluster over time \citep{adams02}. However, these escaped low mass objects may 
remain members of the much larger moving group that surrounds the cluster. 

\citet{jameson08} followed this work by using the wide field camera (WFCAM, \citealt{casali07}) on the United Kingdom Infrared Telescope (UKIRT) to image 
143 known field L dwarfs. These images provided a second epoch for proper motion measurements, when combined with existing 2MASS images, typically taken 7 years 
previously. Using the proper motions and a distance calculated using the spectral type of the L dwarf given by \citet{cruz03}, the moving group method was applied, and all 143 objects 
were scrutinised to check if their direction and magnitude of motion made them candidates of the many moving groups known. 
Members of the Hyades, Ursa Major and Pleiades moving groups were found. 
Radial velocity measurements such as those of \citet{zapatero07} are required however,
before it can be determined if these moving group members  are cluster members that have ``escaped'' as the cluster has dynamically evolved \citep{adams02}. 
Is should also be noted that galactic resonances can produce effects similar to moving groups, and so all members may not be coeval \citep{dehnen98}.

To continue the study started by \citet{bannister07} and \citet{jameson08}, we have measured proper motions for the majority of the 
remaining known field L dwarfs listed in the online L and T dwarf archive (http://spider.ipac.caltech.edu/staff/davy/ARCHIVE/).
This has again been accomplished using the WFCAM on UKIRT and for the more southern objects, Son of ISAAC (SofI) on the 3.58m ESO New Technology Telescope (NTT). 
Using these wide field images and existing catalogue data, we have measured proper motions for an additional 126 L and T dwarfs 
listed in the dwarf archive.

These proper motion data may be put to a number of uses. Using reduced proper motion diagrams they can be used as an approximate measure of distance.
The proper motion measurements can also be used to help identify objects as members of a star cluster or members of a moving group via the moving cluster method.  Taken with measured radial velocities and distances, it can yield all three components 
of velocity (U,V,W).
As brown dwarfs tend to be faint, measuring their radial velocities is very difficult. As a result,  very few L or T dwarfs have measurements, none of which are in this sample. This means we cannot determine whether these moving group members are escaped cluster members or otherwise. 

Our proper motion data are discussed and listed in section 2 of this paper.

\section{Proper Motion Measurements}
\subsection{Data acquisition and reduction}
In order to measure proper motions for known L and T dwarfs we observed a sample of the known L and T dwarfs from the online dwarf archive 
(detailed in Tables \ref{pmsofi} and \ref{pmwfcam}, see http://spider.ipac.caltech.edu/staff/davy/ARCHIVE/ for discovery references) with \textit{J} band 
magnitudes of less than 16.5. 143 of these objects were presented in \citet{jameson08}.
To complete the sample, 126 additional objects have been observed, 88 with declinations between -30$^{\circ}$ and +60$^{\circ}$, 
and 38 with declinations of less than -30$^{\circ}$.
This first group were imaged using WFCAM on UKIRT over the period of June 2006 to March 2008.
WFCAM is a near infrared imager consisting of 4 Rockwell Hawaii-II (HgCdTe 2048x2048) arrays arranged
 such that 4 separately pointed observations can be tiled together to cover a filled square of sky covering 0.75 square degrees 
with 0.4 arcsecond pixels \citep{casali07}. However, as we only required the image of the brown dwarf in question, we only used array 3  
which is regarded as the least noisy array. WFCAM is ideal for this work, as the large field of view per chip means there are many other 
stars in the image, which can be used as astrometric reference stars.
The images were taken in the \textit{J} band in non-photometric conditions using exposure times of $\approx$5-10 minutes and a nine point
 dither pattern. These exposure times gave S/N$\approx$100 even in the poor conditions.

The images were reduced at the Cambridge Astronomical Survey Unit (CASU) using procedures which have been custom written 
for the treatment of WFCAM data (Irwin et al., in preparation, \citealt{dye06}.). In brief, each frame was 
debiased, dark corrected and then flat fielded. The individual dithered images were stacked before having an object detection 
routine run on them. 
The frames were astrometrically calibrated using point sources in the 2MASS catalogue. The accuracy is typically $\approx$0.1'' \citep{dye06} 
The photometric calibration
employed by the CASU pipeline also relies on 2MASS data (there are typically hundreds of 2MASS calibrators per detector) and is found to be
accurate to $\approx$2\% in good conditions \citep{warren07}.  However as we wished to measure proper motions, the astrometric calibration 
was more important than the photometric calibration for these data.

The objects with declinations less than -30$^{\circ}$ were observed using SofI on the NTT. SofI is 
a wide field infrared imager consisting of an Hawaii HgCdTe 1024x1024 array. This has a plate scale  of 0.292 ``/pixel and a field of 
view of $\sim$ 5 arcminutes \citep{moorwood98}. The images were taken between 02/04/2007 and 31/08/2007 in the \textit{J} band. The conditions were 
non-photometric, exposure times were between 5 and 10 minutes, and a nine point dither pattern was used. As for the WFCAM images, 
S/N$\approx$100 was achieved in the conditions.

These data were reduced using the ORAC-DR pipeline \citep{allan02}. This pipeline is maintained and developed by the JAC in Hawaii,
 and has been adapted to reduce SofI data \citep{currie04}.
The pipeline bias corrected, flat fielded, sky subtracted and created mosaics out of the nine images per object.
These mosaics were then astrometrically calibrated using point sources in 2MASS, and the STARLINK Autoastrom package.
 The accuracy of this calibration is typically greater than 0.1'', as measured by the rms of the fit to 
the 2MASS point sources. This is worse than 
the WFCAM astrometry as one might expect, as there are much fewer reference stars in the smaller SofI images.
Object detection was completed using SExtractor  using the image parameters (gain etc) and a suitably sized background mesh, tailored to the seeing and pixel scale of the image.

\subsection{Calculating proper motions}
The astrometry for 2MASS is good to 80 mas over the whole survey, and to 50 mas over a small area \citep{skrutskie06}. 
Because the WFCAM and SofI astrometry is also calibrated to the 2MASS catalogue, accurate relative proper motion measurements could be calculated 
simply by taking the difference in 2MASS and WFCAM positions and dividing by the epoch difference. 
We calculated the epoch difference  by taking the difference in the Julian date as given in the FITS header for each image, which is between
 6.5 and 8.7 years with the average epoch difference being 7.8 years for SofI and between 5.8 and 9.5 with an average of 8.5 for WFCAM.
The proper motion measurements for each object in every WFCAM array 3 and SofI image were calculated by this method. 
This same method was used for our previous 143 objects \citep{jameson08} and produced results consistent with the existing data.

This motion was then converted into mas yr$^{-1}$. The proper motion has been calculated directly from the RA and dec of the object in question, not 
from pixel motion on images, hence $\mu_{\alpha}$=($\Delta\alpha$/$\Delta$T)$\cos\delta$ `` yr$^{-1}$ if $\Delta\alpha$ is converted to arcseconds. 
These proper motions are relative proper motions, in the sense that they are relative to the bulk of the background stars in the field, 
which are generally moving slowly enough to be assumed to have zero motion. 

However, for the WFCAM fields we checked the reference star motion  so that the proper motion of the brown dwarf could be 
altered if there was a standard offset in the field.

The proper motions were separated in $\mu_{\alpha}cos\delta$ and $\mu_{\delta}$ from -500 to 500 mas yr$^{-1}$ in each direction, in bins of 
size 20 mas yr$^{-1}$, and the number of objects 
falling into each bin were totalled. 
We then fitted a two dimensional Gaussian to the data for each field to determine the spread of the reference stars, as well as 
the true centre of the motion. The process was then repeated after the initial fit, rejecting any objects that lay outside 3$\sigma$ 
of the fitted Gaussian,
 before fitting another Gaussian to this data. 
This fitting was important in some cases as the reference stars had quite a large spread, and in other cases the proper motion of our brown dwarf
was of the same order of magnitude as the references.
These centroiding changes, or the centres of the fit, were then subtracted from the calculated proper motion measurements.

We used the $\sigma$ value of the Gaussian to determine the error on our measurements. In general the errors were of the order of $\approx$ 
15 mas yr$^{-1}$. The quoted errors are the $\sigma$ value of 
the Gaussian fitted to the proper motion points for each image. Strictly this should be the $\sigma$ plus the position 
error of the object added in quadrature.
However, 
the centroiding errors for WFCAM are less than 2 mas yr$^{-1}$, and so are small compared to the $\sigma$ value.  

This method could not be used for some of the SofI fields, as there were less than 30 reference stars present, and in some 
cases, as few as 6. For these fields, the Gaussian fitting
 did not converge. As an alternative method of gauging the offset in motion and the proper motion errors, the mean of both 
$\mu_{\alpha}$cos$\delta$ and $\mu_{\delta}$ was calculated for all of the objects in the field, once the brown dwarf had been 
removed. The standard deviation was also calculated, and again an iterative process used to remove any objects that lay further 
from the mean than 3 standard deviations. The process was then repeated and the means were then used as the offset in motion and 
the standard error of the mean values was then used as the error of the proper motion.
The measured proper motions are given in Table \ref{pmsofi} for the brown dwarfs imaged by SofI and in Table \ref{pmwfcam} for the brown dwarfs imaged by WFACM.

\begin{table*}
\caption{\label{pmsofi}2MASS name, RA, Dec, $\mu_{\alpha}$cos$\delta$, $\mu_{\delta}$ and 2MASS magnitudes for all the L and T dwarfs for which we measured proper motions using SofI.}
\begin{center}
\begin{tabular}{l c c c c c c c}
\hline
Name&RA&Dec&$\mu_{\alpha}$cos$\delta$&$\mu_{\delta}$&$J$&$H$&$K_S$\\
2MASS&\multicolumn {2}{|c|}{J2000}&\multicolumn {2}{|c|}{mas yr$^{-1}$}&&&\\
\hline
J00145575-4844171 &  00   14 55.76  &-48   44 17.15&851.29 $\pm$ 12.38 & 279.42$\pm$8.02   & 14.050$\pm$ 0.033 &13.107$\pm$ 0.035& 12.723$\pm$ 0.028\\
J00165953-4056541 &  00   16 59.53  &-40   56 54.15&162.51 $\pm$ 14.44 & 16.06$\pm$5.38    & 15.316$\pm$ 0.060& 14.206$\pm$ 0.047& 13.432$\pm$ 0.037\\
J00325584-4405058 &  00   32 55.84 & -44   05 05.87&116.76 $\pm$ 7.85 & -88.96$\pm$5.45    & 14.776$\pm$ 0.032& 13.857$\pm$ 0.032& 13.269$\pm$ 0.035\\
J00531899-3631102 &  00   53 18.99 & -36   31 10.29&36.82 $\pm$ 22.71 & -72.41$\pm$8.57    & 14.445$\pm$ 0.023& 13.480$\pm$ 0.030& 12.937$\pm$ 0.027\\
J01174748-3403258 &  01   17 47.48 & -34   03 25.82&103.14 $\pm$ 13.98 & -39.70$\pm$7.00   & 15.178$\pm$ 0.034& 14.209$\pm$ 0.038& 13.489$\pm$ 0.036\\
J01253689-3435049 &  01   25 36.90 & -34   35 04.91&120.42 $\pm$ 42.95 & -12.97$\pm$21.05  & 15.522$\pm$ 0.054& 14.474$\pm$ 0.051& 13.898$\pm$ 0.054\\
J01415823-4633574 &  01   41 58.23 & -46   33 57.43&92.90 $\pm$ 10.43 & -5.54$\pm$8.27     & 14.832$\pm$ 0.041& 13.875$\pm$ 0.024& 13.097$\pm$ 0.030\\
J02182913-3133230&   02   18 29.13 & -31   33 23.08&-131.09 $\pm$ 9.87 & -97.18$\pm$16.69  & 14.728$\pm$ 0.038& 13.808$\pm$ 0.036& 13.154$\pm$ 0.033\\
J02550357-4700509 &  02   55 03.58&  -47   00 50.99&1052.88 $\pm$ 11.17 & -546.54$\pm$6.17 & 13.246$\pm$ 0.024& 12.204$\pm$ 0.022 &11.558$\pm$ 0.023\\
J03185403-3421292&   03   18 54.04 & -34   21 29.22&401.63 $\pm$ 9.93 & 43.28$\pm$4.10     & 15.569$\pm$ 0.053& 14.346$\pm$ 0.043& 13.507$\pm$ 0.038\\
J03572695-4417305 &  03   57 26.96 & -44   17 30.55&63.15 $\pm$ 12.71 & 2.35$\pm$7.54      & 14.367$\pm$ 0.029& 13.531$\pm$ 0.025 &12.907$\pm$ 0.026\\
J04430581-3202090 &  04   43 05.81 & -32   02 09.01&-0.46 $\pm$ 13.70 & 205.19$\pm$7.54    & 15.273$\pm$ 0.050& 14.350$\pm$ 0.055& 13.877$\pm$ 0.062\\
J04455387-3048204 &  04   45 53.88 & -30   48 20.46&158.07 $\pm$ 9.33 & -402.39$\pm$5.36   & 13.393 $\pm$0.023& 12.580$\pm$ 0.022 &11.975 $\pm$0.019\\
J04510093-3402150 &  04   51 00.93 & -34   02 15.04&76.34 $\pm$ -8.05 & 166.66$\pm$-8.01   & 13.541$\pm$ 0.020& 12.826$\pm$ 0.022& 12.294$\pm$ 0.024\\
J06244595-4521548 &  06   24 45.95 & -45   21 54.88&-49.71 $\pm$ 37.67 & 392.74$\pm$13.81  & 14.480$\pm$ 0.026& 13.335$\pm$ 0.027& 12.595$\pm$ 0.024\\
J06411840-4322329 &  06   41 18.40 & -43   22 32.93&216.60$\pm$ 15.76 & 642.33$\pm$9.93    & 13.751$\pm$ 0.023& 12.941$\pm$ 0.032& 12.451$\pm$ 0.027\\
J07193188-5051410 &  07   19 31.88 & -50   51 41.06&199.11 $\pm$ 20.49 & -46.44$\pm$13.78  & 14.094$\pm$ 0.029 &13.282$\pm$ 0.033& 12.773$\pm$ 0.026\\
J09221952-8010399 &  09   22 19.52 & -80   10 39.93&3.39 $\pm$ 43.54 & -66.69$\pm$9.36     & 15.276$\pm$ 0.053 &14.285$\pm$ 0.033 &13.681$\pm$ 0.046\\
J10043929-3335189 &  10   04 39.29 & -33   35 18.91&365.99 $\pm$ 25.14 & -350.34$\pm$14.46 & 14.480$\pm$ 0.032& 13.490$\pm$ 0.035& 12.924$\pm$ 0.023\\
J10365305-3441380 &  10   36 53.06 & -34   41 38.09&-32.54 $\pm$ 26.06 & -445.75$\pm$18.42 & 15.622$\pm$ 0.046& 14.446$\pm$ 0.034& 13.798$\pm$ 0.042\\
J11223624-3916054 &  11   22 36.24 & -39   16 05.49&55.74 $\pm$ 15.95 & -171.34$\pm$15.39  & 15.705$\pm$ 0.059& 14.682$\pm$ 0.046& 13.875$\pm$ 0.052\\
J11544223-3400390 &  11   54 42.23 & -34   00 39.06&-158.55 $\pm$ 13.23 & 28.35$\pm$13.07  & 14.195$\pm$ 0.031& 13.331$\pm$ 0.027& 12.851$\pm$ 0.032\\
J12073804-3909050 &  12   07 38.04 & -39   09 05.09&-134.78 $\pm$ 24.57 & 51.10$\pm$12.76  & 14.689$\pm$ 0.038& 13.817$\pm$ 0.026& 13.244$\pm$ 0.037\\
J13411160-3052505 &  13   41 11.60 & -30   52 50.53&35.45 $\pm$ 30.00 & -126.20$\pm$18.70  & 14.607$\pm$ 0.031& 13.725$\pm$ 0.032& 13.081$\pm$ 0.024\\
J13595510-4034582 &  13   59 55.10 & -40   34 58.27&44.60 $\pm$ 17.69 & -491.51$\pm$14.21  & 13.645$\pm$ 0.023& 13.034$\pm$ 0.027& 12.566$\pm$ 0.027\\
J14252798-3650229 &  14   25 27.98 & -36   50 23.00&-253.11 $\pm$ 23.59 & -448.55$\pm$28.18& 13.747$\pm$ 0.026& 12.575$\pm$ 0.020& 11.805$\pm$ 0.025\\
J17534518-6559559 &  17   53 45.18 & -65   59 55.91&-50.42 $\pm$ 78.51 & -329.02$\pm$35.82 & 14.095$\pm$ 0.025& 13.108$\pm$ 0.026& 12.424$\pm$ 0.027\\
J19285196-4356256 &  19   28 51.97 & -43   56 25.64&81.24 $\pm$ 17.23 & -265.82$\pm$25.87  & 15.199$\pm$ 0.042& 14.127$\pm$ 0.043& 13.457$\pm$ 0.036\\
J19360187-5502322 &  19   36 01.88 & -55   02 32.22&210.17 $\pm$ 32.98 & -273.20$\pm$21.09 & 14.486$\pm$ 0.037& 13.628$\pm$ 0.034& 13.046$\pm$ 0.031\\
J20414283-3506442 &  20   41 42.83 & -35   06 44.27&56.16 $\pm$ 15.79 & -118.14$\pm$13.02  & 14.887$\pm$ 0.031& 13.987$\pm$ 0.020& 13.401$\pm$ 0.036\\
J21075409-4544064 &  21   07 54.09 & -45   44 06.47&114.58 $\pm$ 40.06 & -7.85$\pm$30.98   & 14.915$\pm$ 0.029& 13.953$\pm$ 0.037& 13.380$\pm$ 0.033\\
J21420580-3101162 &  21   42 05.80  &-31   01 16.29&45.95 $\pm$ 6.20 & -97.31$\pm$5.09     & 15.844$\pm$ 0.066& 14.767$\pm$ 0.053& 13.965$\pm$ 0.050\\
J21501592-7520367 &  21   50 15.93 & -75   20 36.73&869.96 $\pm$ 22.08 & -277.81$\pm$6.55  & 14.056$\pm$ 0.026& 13.176$\pm$ 0.031& 12.673$\pm$ 0.029\\
J21574904-5534420 &  21   57 49.04 & -55   34 42.05&38.93 $\pm$ 24.51 & -4.58$\pm$12.31    & 14.263$\pm$ 0.029& 13.440$\pm$ 0.028 &13.002$\pm$ 0.029\\
J22064498-4217208 &  22   06 44.98 & -42   17 20.89&140.24 $\pm$ 10.35 & -174.18$\pm$4.41  & 15.555$\pm$ 0.065& 14.447$\pm$ 0.061 &13.609$\pm$ 0.055\\
J23312378-4718274 &  23   31 23.79 & -47   18 27.44&72.07 $\pm$ 40.02 & 0.33$\pm$28.02     & 15.659$\pm$ 0.067& 15.510$\pm$ 0.149& 15.389$\pm$ 0.196\\
\hline
\end{tabular}
\end{center}
\end{table*}

\begin{table*}
\caption{\label{pmwfcam}2MASS name, 2MASS RA, 2MASS Dec, $\mu_{\alpha}$cos$\delta$, $\mu_{\delta}$ and 2MASS magnitudes for all the L and T dwarfs for which we measured proper motions using WFCAM.}
\begin{center}
\begin{tabular}{l c c c c c c c}
\hline
Name&RA&Dec&$\mu_{\alpha}$cos$\delta$&$\mu_{\delta}$&$J$&$H$&$K_S$\\
2MASS&\multicolumn {2}{|c|}{J2000}&\multicolumn {2}{|c|}{mas yr$^{-1}$}&&&\\
\hline
J00135779-2235200&    00   13 57.80 & -22   35 20.09& 57.74$\pm$ 23.76 & -60.74$\pm$ 21.38& 15.775$\pm$ 0.064& 14.595$\pm$ 0.062 &14.036$\pm$ 0.050\\
J00332386-1521309&    00   33 23.86 & -15   21 30.94 &330.31$\pm$ 16.25&  46.33 $\pm$22.26& 15.286$\pm$ 0.055 &14.208$\pm$ 0.051& 13.410$\pm$ 0.038\\
J00345684-0706013&    00   34 56.84 &  -07    06  01.32 &215.35 $\pm$21.32&  -138.55$\pm$ 15.29 &15.531$\pm$ 0.059& 14.566$\pm$ 0.041& 13.942$\pm$ 0.065\\
J00511078-1544169&    00   51 10.79 & -15   44 16.91 &76.37$\pm$ 24.08 & -4.25$\pm$ 19.61 &15.277$\pm$ 0.049 &14.164$\pm$ 0.047& 13.466$\pm$ 0.038\\
J00584253-0651239&    00   58 42.53 &  -06   51 23.94 &158.29$\pm$ 20.18&  -105.96 $\pm$15.76 &14.311$\pm$ 0.023& 13.444$\pm$ 0.028& 12.904$\pm$ 0.032\\
J01311838+3801554&    01   31 18.39 &  38    01 55.48 &393.88$\pm$ 15.45&  -21.70$\pm$ 13.25 &14.679$\pm$ 0.032& 13.696$\pm$ 0.033& 13.054$\pm$ 0.033\\
J01353586+1205216&    01   35 35.86 &  12    05 21.67 &-45.23$\pm$ 14.82&  -441.77$\pm$ 14.08 &14.412$\pm$ 0.030& 13.527$\pm$ 0.031& 12.918$\pm$ 0.028\\
J01410321+1804502&    01   41  03.22&   18    04 50.20& 425.05$\pm$ 17.35&  -32.16$\pm$ 16.50 &13.875$\pm$ 0.022& 13.034$\pm$ 0.024& 12.492$\pm$ 0.026\\
J01443536-0716142&    01   44 35.36 &  -07   16 14.23& 408.28$\pm$ 17.56&  -187.65$\pm$ 17.99 &14.191$\pm$ 0.023& 13.008$\pm$ 0.027& 12.268$\pm$ 0.021\\
J01473344+3453112&    01   47 33.45 &  34   53 11.24 &47.24$\pm$ 18.30 & -47.19$\pm$ 14.84 &14.946$\pm$ 0.037 &14.162$\pm$ 0.040 &13.574 $\pm$0.037\\
J02050344+1251422&    02    05  03.44 &  12   51 42.23 &388.84$\pm$ 12.17&  -8.71$\pm$ 12.15& 15.679$\pm$ 0.055 &14.449$\pm$ 0.046& 13.671$\pm$ 0.035\\
J02073557+1355564&    02    07 35.57 &  13   55 56.45 &259.66$\pm$ 12.71&  -159.71$\pm$ 13.59 &15.462$\pm$ 0.048& 14.474$\pm$ 0.043& 13.808$\pm$ 0.045\\
J02085499+2500488&    02    08 55.00 &  25    00 48.82 &0.71 $\pm$15.49&  25.63$\pm$ 11.047 &16.206$\pm$ 0.092& 14.974$\pm$ 0.080& 14.405$\pm$ 0.069\\
J02081833+2542533&    02    08 18.33 &  25   42 53.31 &386.59$\pm$ 14.268&  -25.14$\pm$ 16.73 &13.989$\pm$ 0.023 &13.107$\pm$ 0.029& 12.588$\pm$ 0.025\\
J02082363+2737400&    02    08 23.64 &  27   37 40.09 &231.06$\pm$ 11.62&  -95.82$\pm$ 16.35 &15.714$\pm$ 0.059& 14.560$\pm$ 0.059& 13.872$\pm$ 0.051\\
J02112827+1410039&    02   11 28.28 &  14   10  03.95 &-66.39$\pm$ 13.83&  -27.22$\pm$ 13.39 &16.128$\pm$ 0.077 &15.423 $\pm$0.089& 15.009$\pm$ 0.123\\
J02132880+4444453 &   02   13 28.80 &  44   44 45.36& -42.20$\pm$ 13.86&  -132.39$\pm$ 13.23 &13.494$\pm$ 0.021 &12.757 $\pm$0.023& 12.213$\pm$ 0.021\\
J02271036-1624479&    02   27 10.36 & -16   24 47.95& 438.00$\pm$ 18.91&  -286.59$\pm$ 18.81 &13.573$\pm$ 0.020 &12.630$\pm$ 0.022 &12.143$\pm$ 0.029\\
J02301551+2704061&    02   30 15.51 &  27    04  06.18 &198.10 $\pm$15.38&  -13.50$\pm$ 13.63 &14.294$\pm$ 0.024& 13.478$\pm$ 0.028& 12.986$\pm$ 0.021\\
J02354756-0849198&    02   35 47.57 &  -08   49 19.80 &-12.28 $\pm$16.75&  21.33$\pm$ 16.51& 15.571$\pm$ 0.054& 14.812$\pm$ 0.055& 14.191$\pm$ 0.066\\
J02361794+0048548&    02   36 17.94 &   00   48 54.82 &161.33 $\pm$10.10 & -176.33$\pm$ 19.16 &16.098$\pm$ 0.076& 15.265$\pm$ 0.066 &14.666$\pm$ 0.090\\
J02394245-1735471&    02   39 42.46 & -17   35 47.20 &61.09$\pm$13.80 & -79.12$\pm$ 18.79& 14.291$\pm$ 0.029& 13.525$\pm$ 0.034& 13.039$\pm$ 0.030\\
J02411151-0326587&    02   41 11.52 &  -03   26 58.78 &93.43 $\pm$17.00&  -19.87$\pm$ 13.40 &15.799$\pm$ 0.064& 14.811$\pm$ 0.053 &14.035$\pm$ 0.049\\
J02424355+1607392&    02   42 43.55 &  16    07 39.27 &163.62$\pm$ 14.16&  -198.32$\pm$ 12.77 &15.776$\pm$ 0.052& 14.998$\pm$ 0.053& 14.349$\pm$ 0.057\\
J02560189+0110467&    02   56  01.89&    01   10 46.71 &66.48$\pm$ 22.28&  -62.38$\pm$ 22.06& 16.212$\pm$ 0.102& 15.696$\pm$ 0.184 &15.216$\pm$ 0.175\\
J03090888-1949387&    03    09  08.89 & -19   49 38.76 &218.00$\pm$ 11.57&  -13.33$\pm$ 11.90 &15.752$\pm$ 0.055& 14.656$\pm$ 0.062& 14.057 $\pm$0.065\\
J03101401-2756452 &   03   10 14.01 & -27   56 45.27 &-118.89$\pm$ 17.61 & -46.70$\pm$ 16.21 &15.795$\pm$ 0.070& 14.662$\pm$ 0.049 &13.959$\pm$ 0.060\\
J03105986+1648155&	03   11  00.02 &  16   48 15.68 &	262.83 $\pm$ 19.28& 9.38$\pm$ 17.90&	 16.025$\pm$   0.083& 14.932$\pm$   0.070& 14.312 $\pm$  0.067\\   
J03140344+1603056&    03   14  03.45&   16    03  05.63 &-244.73$\pm$ 13.31 & -49.07$\pm$ 13.91& 12.526$\pm$ 0.021& 11.823$\pm$ 0.035& 11.238 $\pm$0.019\\
J03164512-2848521&    03   16 45.13 & -28   48 52.17 &105.25$\pm$ 22.15&  -80.97$\pm$2 12.88& 14.578$\pm$ 0.039& 13.772$\pm$ 0.035& 13.114$\pm$ 0.035\\
J03202839-0446358&	03   20 28.24 &  -04   46 40.59 &	-249.91 $\pm$ 17.00 &-501.77$\pm$ 12.26&	 13.259$\pm$   0.021& 12.535$\pm$   0.022& 12.134 $\pm$  0.024 \\   
J03264225-2102057&    03   26 42.26&  -21    02  05.77 &103.12$\pm$ 16.14 & -129.87$\pm$ 18.72& 16.134 $\pm$0.093 &14.793$\pm$ 0.075& 13.922$\pm$ 0.065\\
J03281738+0032572&	03   28 17.49 &   00   32 57.57 &	201.84 $\pm$ 22.19& 35.50$\pm$ 17.21&	 15.988$\pm$   0.091& 14.975$\pm$   0.074& 14.161 $\pm$  0.077 \\   
J03301774+0000477&	03   30 17.74 &   00    00 47.44 &	-8.29 $\pm$ 11.88& -33.00$\pm$ 11.08&	 16.520$\pm$   0.111& 15.881$\pm$   0.130& 15.525 $\pm$  0.182\\    
J03504861-0518126&    03   50 48.61&   -05   18 12.70 &19.91$\pm$ 11.31&  -8.04$\pm$ 15.59 &16.327$\pm$ 0.092& 15.525$\pm$ 0.094& 15.125 $\pm$0.134\\
J03552337+1133437&    03   55 23.37 &  11   33 43.71& 223.16$\pm$ 22.90 & -607.16$\pm$ 22.07 &14.050$\pm$ 0.020& 12.530$\pm$ 0.029& 11.526$\pm$ 0.019\\
J03554191+2257016 &   03   55 41.91 &  22   57  01.68 &171.90$\pm$ 13.14&  -21.38$\pm$ 10.33& 16.111$\pm$ 0.077& 15.052$\pm$ 0.068 &14.284$\pm$ 0.060\\
J03572110-0641260&    03   57 21.10 &  -06   41 26.04 &133.59$\pm$ 22.58&  27.12$\pm$5 25.51& 15.953$\pm$ 0.081& 15.060$\pm$ 0.085& 14.599$\pm$ 0.089\\
J04070885+1514565&    04    07  08.85 &  15   14 56.60 &97.97$\pm$ 16.28&  -74.65$\pm$ 17.40& 16.055$\pm$ 0.091& 16.017$\pm$ 0.208& 15.922$\pm$ 0.261\\
J04082905-1450334&	04    08 29.18&  -14   50 34.33 &	193.43 $\pm$ 11.35& -99.07$\pm$ 17.75&	 14.222$\pm$   0.028& 13.337$\pm$   0.029& 12.817 $\pm$  0.021\\    
J04090950+2104393&    04    09  09.51&   21    04 39.38 &101.13$\pm$ 15.14&  -148.46$\pm$ 12.25& 15.508$\pm$ 0.053& 14.497$\pm$ 0.045& 13.850$\pm$ 0.046\\
J04132039-0114248&    04   13 20.39  & -01   14 24.86& 61.22$\pm$ 15.29 & 0.08$\pm$ 18.67 &15.303$\pm$ 0.047& 14.657 $\pm$0.033& 14.135$\pm$ 0.059\\
J04285096-2253227&	04   28 51.05 & -22   53 21.0 &	123.44 $\pm$ 16.02& 183.48$\pm$ 15.70&	 13.507 $\pm$  0.020 &12.668$\pm$   0.026& 12.118 $\pm$  0.024   \\ 
J04390101-2353083&	04   39  00.93&  -23   53  09.51 &	-119.29 $\pm$ 22.66&-127.25$\pm$ 15.05&	 14.408$\pm$   0.026& 13.409$\pm$   0.027& 12.816  $\pm$ 0.021 \\   
J04532647-1751543&	04   53 26.51 & -17   51 54.35 &	50.80$\pm$ 14.38& -4.14$\pm$ 14.30&	 15.142$\pm$   0.033& 14.060$\pm$   0.034& 13.466 $\pm$  0.034 \\   
J05012406-0010452&	05    01 24.19&    00   10 46.58 &	192.82$\pm$ 11.98& -139.86$\pm$ 12.29&	 14.982$\pm$   0.036& 13.713$\pm$   0.033& 12.963 $\pm$  0.034 \\   
J05021345+1442367&    05    02 13.45 &  14   42 36.78 &71.55$\pm$ 13.68&  -1.04$\pm$ 18.53 &14.271$\pm$ 0.022& 13.392$\pm$ 0.019& 12.955$\pm$ 0.028\\
J05120636-2949540&	05   12  06.36&  -29   49 53.12 &	-6.40$\pm$ 13.41& 96.87$\pm$ 12.28&	 15.463$\pm$   0.055& 14.156$\pm$   0.047& 13.285 $\pm$  0.041 \\   
J05184616-2756457&	05   18 46.19 & -27   56 45.69 &	37.27$\pm$ 17.01& 10.72$\pm$ 19.72&	 15.262$\pm$   0.041& 14.295$\pm$   0.045& 13.615  $\pm$ 0.039\\    
J05233822-1403022 &   05   23 38.22 & -14    03  02.29 &109.87$\pm$ 17.14&  178.88$\pm$ 21.12 &13.084$\pm$ 0.021& 12.220 $\pm$0.020& 11.638 $\pm$0.025\\
J06050196-2342270&    06    05  01.96 & -23   42 27.01& -66.48$\pm$ 15.90&  117.68$\pm$ 18.21 &14.512$\pm$ 0.032& 13.727 $\pm$0.035& 13.145$\pm$ 0.030\\
J06262121+0029341&    06   26 21.21 &   00   29 34.14 &47.15$\pm$ 12.94&  1.03$\pm$ 12.78 &15.925$\pm$ 0.093& 15.209$\pm$ 0.120& 14.860$\pm$ 0.123\\
J06523073+4710348 &   06   52 30.74 &  47   10 34.83 &-129.78$\pm$ 14.67&  149.52$\pm$7 12.78& 13.511$\pm$ 0.020& 12.384$\pm$ 0.023& 11.694 $\pm$0.018\\
J07171626+5705430&    07   17 16.27 &  57    05 43.05 &-17.99$\pm$ 17.87&  67.17$\pm$ 15.13& 14.636$\pm$ 0.029& 13.593$\pm$ 0.028& 12.945$\pm$ 0.025\\
J07231462+5727081&    07   23 14.62 &  57   27  08.17& 61.39$\pm$ 22.71&  -219.85$\pm$ 12.31 &13.970$\pm$ 0.024& 13.156$\pm$ 0.028& 12.613$\pm$ 0.029\\
J07400966+3212032&    07   40  09.67&   32   12  03.24 &-31.98$\pm$ 11.79&  -80.79$\pm$ 13.45& 16.191$\pm$ 0.090& 14.862$\pm$ 0.059& 14.222$\pm$ 0.059\\
J07414920+2351275 &   07   41 49.21 &  23   51 27.51& -250.22$\pm$ 12.18&  -116.21$\pm$ 13.32& 16.158$\pm$ 0.100& 15.838$\pm$ 0.185& 15.847$\pm$ 99.99\\
J07420130+2055198 &   07   42  01.30 &  20   55 19.88 &-318.67$\pm$ 12.08&  -229.57$\pm$ 13.15 &16.193 $\pm$0.090& 15.911$\pm$ 0.181& 15.225$\pm$ 99.99\\
J07475631+3947329&    07   47 56.31 &  39   47 32.92 &55.04$\pm$ 14.43 & -41.72$\pm$ 13.26 &15.076$\pm$ 0.039& 14.163$\pm$ 0.038& 13.724$\pm$ 0.044\\
J07525942+4136344&    07   52 59.43&   41   36 34.49 &0.00$\pm$ 13.39&  43.98$\pm$ 10.46& 16.356$\pm$ 0.130 &15.601$\pm$ 0.160 &15.191$\pm$ 99.99\\

\hline
\end{tabular}
\end{center}
\end{table*}
\addtocounter{table}{-1}
 \begin{table*}
\caption{continued}
\begin{center}
\begin{tabular}{l c c c c c c c}
\hline
Name&RA&Dec&$\mu_{\alpha}$cos$\delta$&$\mu_{\delta}$&$J$&$H$&$K_S$\\
2MASS&\multicolumn {2}{|c|}{J2000}&\multicolumn {2}{|c|}{mas yr$^{-1}$}&&&\\
\hline
J07533217+2917119&    07   53 32.17 &  29   17 11.93 &-88.43$\pm$ 11.08&  -85.34$\pm$ 13.12& 15.516$\pm$ 0.046& 14.527$\pm$ 0.039& 13.849$\pm$ 0.042\\
J07554795+2212169&    07   55 47.95 &  22   12 16.94 &-9.87$\pm$ 14.88&  -226.48$\pm$ 17.78& 15.728$\pm$ 0.063& 15.669$\pm$ 0.144& 15.753$\pm$ 0.207\\
J07584037+3247245&    07   58 40.37 &  32   47 24.55 &-204.23$\pm$ 18.01&  -316.21$\pm$ 12.42& 14.947$\pm$ 0.043& 14.111$\pm$ 0.041& 13.879$\pm$ 0.056\\
J08014056+4628498&    08    01 40.56&   46   28 49.84& -194.73$\pm$ 23.54&  -331.056$\pm$ 19.10& 16.275$\pm$ 0.133 &15.452$\pm$ 0.142 &14.536$\pm$ 0.100\\
J08053189+4812330&    08    05 31.89&   48   12 33.10& -455.48$\pm$ 14.62&  61.35$\pm$ 16.32& 14.734$\pm$ 0.034& 13.917$\pm$ 0.040& 13.444$\pm$ 0.040\\
J08155674+4524119&    08   15 56.75 &  45   24 11.93 &-31.49$\pm$ 15.56 & -42.96$\pm$13.18& 16.057$\pm$ 0.078 &15.233$\pm$ 0.093& 14.812$\pm$ 0.097\\
J08202996+4500315&    08   20 29.96 &  45    00 31.52& -104.16$\pm$ 25.82&  -299.16$\pm$14.76& 16.279$\pm$ 0.107& 15.000$\pm$ 0.086& 14.218$\pm$ 0.065\\
J08234818+2428577&    08   23 48.18 &  24   28 57.71 &-160.30$\pm$ 19.81&  73.19$\pm$ 19.17 &14.986$\pm$ 0.042& 14.060$\pm$ 0.044& 13.377$\pm$ 0.029\\
J08290664+1456225 &   08   29  06.64&   14   56 22.56 &-49.41$\pm$ 16.67 & -227.11$\pm$ 18.47 &14.750$\pm$ 0.028& 13.801$\pm$ 0.035& 13.166$\pm$ 0.031\\
J08304878+0128311&    08   30 48.78 &   01   28 31.15 &222.11$\pm$ 15.89 & -310.54$\pm$ 15.36 &16.289$\pm$ 0.111& 16.140$\pm$ 0.213 &16.358$\pm$ 99.99\\
J08320451-0128360&    08   32  04.52&   -01   28 36.05& 63.83$\pm$ 13.19&  26.73$\pm$ 15.10 &14.128$\pm$ 0.028 &13.318$\pm$ 0.022& 12.712$\pm$ 0.026\\
J08355829+0548308&    08   35 58.30&    05   48 30.85 &-109.70$\pm$ 12.79&  -15.00$\pm$ 15.54 &14.533$\pm$ 0.034& 13.683$\pm$ 0.036& 13.168$\pm$ 0.033\\
J08472872-1532372&    08   47 28.73&  -15   32 37.21 &149.07$\pm$ 16.12&  -177.78 $\pm$15.82& 13.513$\pm$ 0.023 &12.629$\pm$ 0.026& 12.061$\pm$ 0.021\\
J08523490+4720359&    08   52 34.91&   47   20 35.91 &-28.69 $\pm$18.13 & -386.53$\pm$ 13.88 &16.182$\pm$ 0.108& 15.419$\pm$ 0.146& 14.718$\pm$ 0.116\\
J08564793+2235182&    08   56 47.94 &  22   35 18.21 &-184.24$\pm$ 19.74 & 3.01$\pm$ 20.71 &15.679 $\pm$0.064 &14.580$\pm$ 0.052& 13.951$\pm$ 0.046\\
J08575849+5708514&    08   57 58.49 &  57    08 51.42 &-413.61$\pm$ 20.52 & -353.43$\pm$16.85& 15.038$\pm$ 0.038& 13.790$\pm$ 0.041& 12.962 $\pm$0.028\\
J08592547-1949268&    08   59 25.48&  -19   49 26.89 &-310.77 $\pm$13.51 & -78.59$\pm$15.95& 15.527$\pm$ 0.052 &14.436$\pm$ 0.041& 13.751$\pm$ 0.057\\
J09095749-0658186 &   09    09 57.49 &  -06   58 18.64 &-174.53$\pm$ 14.89&  39.70$\pm$ 16.48 &13.890$\pm$ 0.021& 13.090$\pm$ 0.020& 12.539$\pm$ 0.024\\
J09153413+0422045 &   09   15 34.14 &   04   22  04.59& -88.46$\pm$ 18.02&  40.27$\pm$ 20.60& 14.548 $\pm$0.028& 13.531$\pm$ 0.031& 13.011$\pm$ 0.041\\
J09183815+2134058&    09   18 38.16 &  21   34  05.82 &353.41$\pm$ 14.50 & -454.02$\pm$ 15.66 &15.662$\pm$ 0.060& 14.580$\pm$ 99.99& 13.903$\pm$ 0.042\\
J09201223+3517429&    09   20 12.23&   35   17 42.97 &-172.34$\pm$ 13.35 & -185.26$\pm$ 12.72& 15.625$\pm$ 0.062 &14.673$\pm$ 0.056& 13.979$\pm$ 0.061\\
J09211410-2104446&    09   21 14.11&  -21    04 44.60 &260.71$\pm$ 15.59 & -900.91$\pm$ 13.20& 12.779$\pm$ 0.021& 12.152$\pm$ 0.020& 11.690$\pm$ 0.021\\
J09283972-1603128&    09   28 39.72&  -16    03 12.86& -132.85$\pm$15.59&  36.00$\pm$ 13.20& 15.322 $\pm$0.041& 14.292$\pm$ 0.036& 13.615$\pm$ 0.050\\
J09352803-2934596&    09   35 28.04&  -29   34 59.62& 8.84$\pm$ 18.10 & 86.92$\pm$14.61 &14.036$\pm$ 0.026& 13.312$\pm$ 0.027& 12.822$\pm$ 0.026\\
J10185879-2909535&   10   18 58.79 & -29    09 53.56 &-323.01$\pm$ 21.04  &-82.84$\pm$ 15.20 &14.213$\pm$ 0.028& 13.418$\pm$ 0.022& 12.796$\pm$ 0.021\\
J10432508+0001482&	10   43 25.00 &   00    01 46.92 &	-159.31$\pm$ 15.56& -143.85$\pm$ 21.27&	 15.935 $\pm$  0.080& 15.208 $\pm$  0.072& 14.472  $\pm$ 0.100\\    
J12285538+0050440&	12   28 55.37 &   00   50 43.94 &	-32.33$\pm$3.54& -12.41$\pm$2.06&	 15.613$\pm$   0.060& 14.825$\pm$   0.059& 14.162 $\pm$  0.077\\  
J16360078-0034525&	16   36  00.59 &   00   34 54.36 &	-344.71$\pm$ 14.13& -200.89$\pm$ 14.86&	 14.590 $\pm$  0.043& 13.904$\pm$   0.042& 13.415 $\pm$  0.035 \\   
J17434148+2127069&	17   43 41.59&   21   27  09.16 &	165.91$\pm$ 14.890& 248.68$\pm$ 14.47& 15.830$\pm$   0.088& 14.785$\pm$   0.064& 14.321 $\pm$  0.097   \\ 

\hline
\end{tabular}
\end{center}
\end{table*}

Some of the L dwarfs have had their proper motions measured by other people namely \citet{caballero07} and \citet{schmidt07}.
The proper motions given by \citet{caballero07} are taken from the SuperCOSMOS Sky Survey.
The errors on the proper motions are typically 10 mas yr$^{-1}$, although in some cases this is larger, such as in the case of  J0847-15 and J0909-07.
This is because the SuperCOSMOS sky survey which consists of scanned sky atlas photographic plates, imaged in at least two epochs \citep{hambly01a, hambly01c}, has necessarily measured the proper motions over an unfavourable (i.e. short) epoch difference (using the minimum of 2 plates required to calculate the motions).
Both of these objects have \citet{schmidt07} proper motions that do agree with our values, and so the \citet{caballero07} measurements 
are taken to be inaccurate.
Tables \ref{scmidt} and \ref{caballero} show the proper motion comparisons as given by \citet{caballero07} and \citet{schmidt07}.

\begin{table*}
\caption{\label{scmidt}Name,   $\mu$ from \citet{schmidt07}, $\theta$ from \citet{schmidt07}, $\mu$, $\theta$ for all the L and T dwarfs common with \citet{schmidt07}.}
\begin{center}
\begin{tabular}{l c c c c }
\hline
Name&$\mu$ (Schmidt)&$\theta$ (Schmidt)&$\mu$&$\theta$\\
2MASS&'' yr$^{-1}$&$^{\circ}$&'' yr$^{-1}$&$^{\circ}$\\
\hline
J0213+44&0.17$\pm$0.06&195$\pm$21&0.138$\pm$0.01&197$\pm$6\\
J0255-47&1.23$\pm$0.11&120$\pm$5&1.18$\pm$0.01&117$\pm$1\\
J0314+16&0.25$\pm$0.07&254$\pm$13&0.25$\pm$0.01&256$\pm$3\\
J0355+11&0.70$\pm$0.05&159$\pm$3&0.64$\pm$0.02&160$\pm$2\\
J0439-23&0.20$\pm$0.17&220$\pm$37&0.17$\pm$0.02&223$\pm$6\\
J0445-30&0.38$\pm$0.11&161$\pm$23&0.43$\pm$0.01&160$\pm$1\\
J0523-14&0.08$\pm$0.16&21$\pm$105&0.2$\pm$0.02&32$\pm$5\\
J0624-45&0.28$\pm$0.19&19$\pm$52&0.39$\pm$0.01&352$\pm$5\\
J0651+47&0.14$\pm$0.05&336$\pm$13&0.19$\pm$0.013&319$\pm$4\\
J0847-15&0.27$\pm$0.05&146$\pm$10&0.23$\pm$0.015&140$\pm$4\\
J0915+04&0.14$\pm$0.15&284$\pm$84&0.097$\pm$0.018&294$\pm$11\\
J0921-21&0.98$\pm$0.15&163$\pm$10&0.93$\pm$0.113&164$\pm$1\\
J1425-36&0.53$\pm$0.10&204$\pm$7&0.52$\pm$0.03&209$\pm$3\\
J1753-65&0.36$\pm$0.09&178$\pm$24&0.33$\pm$0.04&189$\pm$13\\
\hline
\end{tabular}
\end{center}
\end{table*}
It should be pointed out that 0915+04 is described as a binary by \citet{reid06b} with a separation of 0.73" at -155.0$^{\circ}$. 
Our image also shows two resolved objects which may be why there is such a difference between the \citet{schmidt07} measurements and these data. 

\begin{table*}
\caption{\label{caballero}Name, $\mu$ from \citet{caballero07}, $\theta$ from \citet{caballero07}, $\mu$, $\theta$ for all the L and T dwarfs common with \citet{caballero07}.}
\begin{center}
\begin{tabular}{l c c c c }
\hline
Name&$\mu_\alpha$cos$\delta$ (Caballero)&$\mu_\delta$(Caballero)&$\mu_\alpha$cos$\delta$&$\mu_\delta$\\
2MASS&\multicolumn {4}{|c|}{mas yr$^{-1}$}\\
\hline
J0255-47&1060$\pm$50&-630$\pm$50&1052.88 $\pm$ 11.17 & -546.54$\pm$6.17 \\
J0445-30&167$\pm$12&-424$\pm$12&158.07 $\pm$ 9.33 & -402.39$\pm$5.36 \\
J0451-34&94$\pm$17&114$\pm$16&76.34 $\pm$ -8.05 & 166.66$\pm$-8.01\\
J0719-50&140$\pm$30&-10$\pm$30&199.11 $\pm$ 20.49 & -46.44$\pm$13.78 \\
J0847-15&-130$\pm$160&-20$\pm$180&149.07$\pm$ 16.12&  -177.78 $\pm$15.82\\
J0909-07&-280$\pm$190&110$\pm$180&-174.53$\pm$ 14.89&  39.70$\pm$ 16.48\\
J0921-21&100$\pm$60&-900$\pm$60&260.71$\pm$ 15.59 & -900.91$\pm$ 13.20\\
\hline
\end{tabular}
\end{center}
\end{table*}

Two further two objects have additional proper motion measurements.
J0255-47 has a proper motion measurement of 1.14$\pm$0.0022'' yr$^{-1}$ and a $\theta$ measurement of 119.5$\pm$0.21$^{\circ}$ \citep{costa06}. 
Our measurements are $\mu$=1.18$\pm$0.01 `` yr$^{-1}$ and $\theta$=117.4$\pm$1 $^{\circ}$. This object also has an additional parallax measurement of $\sim$ 5 pc \citep{costa06}, placing it well
within the local neighbourhood.

J0320-04 is suspected to be an unresolved M8.5+T5 binary \citep{burgasser08} and has a proper motion of 0.562$\pm$0.005, and a $\theta$ of 205.9, 
which compares favourably with our measurements of 
0.65$\pm$0.01'' and 206.5$\pm$2 degrees.

\section*{Possible Binaries}
This study also provides an opportunity to search for wide common proper motion companions to the known brown dwarfs. 
Any objects within the field that have a proper motion within 15 mas yr$^{-1}$ of the dwarf's motion were considered as possible companions.
Obviously any brown dwarf with a proper motion close to zero, or the majority of background sources, will have a large number of ``companions'' within our selection.
These objects have been excluded.

Two brown dwarfs J100-33 and J0147+34 appear to have common proper motion companions. 

J1004-33 has a possible M dwarf companion which SIMBAD names  as LHS5166, a high proper motion star with proper motion of $\mu_{\alpha}$cos$\delta$=400 mas yr$^{-1}$ and $\mu_{\delta}$=-420 mas yr$^{-1}$.
Our measurement of its proper motion is 370.66$\pm$25.14, -342.62$\pm$14.46 mas yr$^{-1}$.

J0147+34 has a companion which is an M dwarf as identified by \citet{wei99}. This M dwarf is an X-ray source (2E 0144.7+3438 as given by SIMBAD), suggesting that it may be young.
It is ~43'' away from the known brown dwarf and has a proper motion of 46.2$\pm$18.3, -46.4$\pm$14.8 mas yr$^{-1}$.

Further proper motion and radial velocity measurements are required before these can be confirmed as true companions however.

\section*{Fast moving objects}
In paper I \citep{jameson08}, we discovered 8 high velocity L dwarfs which we suggested may belong tho the thick disk or halo population.
 Applying the same criteria to these data (that the proper motion must
 be greater than 0.85'' yr$^{-1}$), we found 3 candidates from the SofI images. The distance estimate derived from their spectral type as 
defined in \citet{cruz03} is 22.35 pc for J0014-48 (L2), 5.0 pc for J0255-47 (L8) and 25.8 pc for J2150-75 (L1).
J0255-47, also has a parallax measurement from \citet{costa06}, which is 4.96 pc, in agreement with the spectral type distance. 
This object is relatively nearby, which gives it a true velocity of only 28 km s$^{-1}$. The other two objects have velocities of 95 km s$^{-1}$ and 
112 km s$^{-1}$ respectively. 
Both J0014-48 and J2150-75 have relatively blue $J-K_{S}$ and $H-K_{S}$ colours for their spectral types, as with other high velocity dwarfs
 \citep{jameson08,schmidt07}, indicating that they too may be members of an older ($\sim$ 10 Gyr), thick disk population.

From the WFCAM data, just one object met the required selection criteria, J0921-21, 
an L2 dwarf, with a spectral type distance of 12.4 pc, and a velocity of 55 km s$^{-1}$. This velocity is not high 
enough for this object to be considered a true high velocity dwarf.

\section*{Moving groups}
We have attempted to determine if any of these brown dwarfs are members of known moving groups. Being a moving group member means that an age can be estimated for the dwarf. This does not necessarily mean that the moving group member is an ``escaped'' cluster member however, as galactic resonances can produce a similar effect \citep{dehnen98}.
To determine if any of these field dwarf are moving group members, we used the moving cluster method as described in \citet{bannister07} and \citet{jameson08}.
Again, we used the proper motion measurements and direction of that  motion as well as a distance derived from the spectral type as described in \citet{cruz03}.
If the difference in angles between the measured and predicted motion towards the convergent point is less than 14$^{\circ}$ plus the error on the measurement, and the ratio between the 
moving cluster distance and the Cruz spectral type distance was greater than 0.72 and less than 1.28 as in \citet{bannister07} and \citet{jameson08}, then the selected objects were plotted on a colour$-$magnitude diagram, and compared to the DUSTY and NextGen models \citep{chabrier00,baraffe98} and to empirical isochrones defined by known cluster members (\citealt{casewell07} for the Pleiades and \citealt{hogan08} for the Hyades.)

\subsection*{Hyades}
The Hyades cluster has a distance of 46 pc and covers $\approx$ 20$^{\circ}$ of the sky. The Hyades has, until recently, been thought to contain almost no
 low mass members. Extensive searches such as those of \citet{gizis99} and \citet{dobbie02} turned up no brown dwarf members. It was hypothesised that being an older open star cluster (625 Myr; \citealt{perryman98}) any low mass members would have evaporated from the cluster through dynamical evolutionary processes. For a cluster of this age $\sim$ 70\% of stars and $\sim$ 85\% of brown dwarfs are expected to have escaped the cluster \citep{adams02}. 

Recently however, studies using deep, wide field surveys such as the UKIRT Deep Infrared Sky Survey \citep{hogan08} have unearthed 12 L dwarf candidate 
members. \citet{bouvier08} have also claimed two T dwarf members from a 16 square degree survey of the cluster's centre.

\citet{chereul98} first identified escaped Hyads, and more 
recently \citet{bannister07} have identified 7 L and T field dwarfs that belong to the Hyades moving group. \citet{zapatero07} have confirmed that
 one of these objects (2MASS J1217110-031113) has a radial velocity consistent with being an escaped member of the Hyades cluster, while two (2MASS J0205293-115930 and 2MASS J16241436+0029158) have radial velocities that are only consistent with being moving group members. 

In paper I \citep{jameson08}, we reported the discovery of 15 new moving group candidate members.
The Hyades moving group has its convergent point situated at $\alpha$=6$^{h}$29.48$^{m}$, $\delta$=-6$^{\circ}$53.4', and the members have a 
space velocity of 46 km s$^{-1}$ (Madsen, Dravins \& Lindgren, 2002). 

\begin{figure*}
\begin{center}
\scalebox{0.50}{\includegraphics[angle=270]{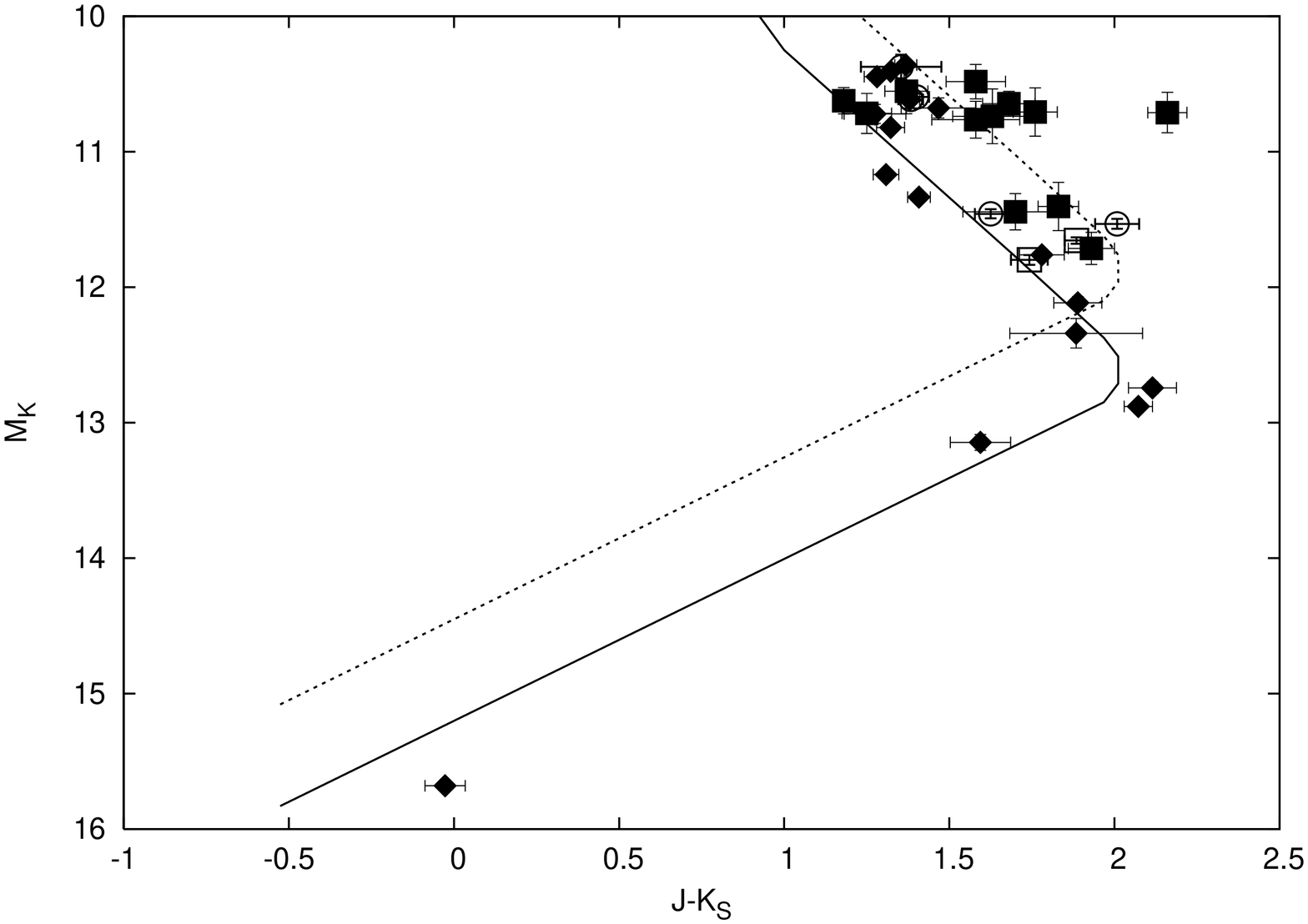}}
\caption{
\label{mg_hyades}
M$_{\rm K}$, $J$-$K$ colour magnitude diagram for the Hyades moving group. The cluster members members identified by \citet{hogan08} are plotted as filled squares.  All of the objects that were previously selected as moving group members are marked as filled diamonds \citep{jameson08, bannister07}. The selected members are marked by a box for objects from the SofI data and by circles for the WFCAM objects. The errors are poissonion and from the photometry only. The solid and dashed lines represent the single star (solid line) and binary (dotted line) cluster sequences as shown in \citet{hogan08}.}
\end{center}
\end{figure*}

After using the moving group method, the \citet{cruz03} distance and the convergent point from \citet{madsen02}, we have found 7 new candidate 
members of the Hyades moving group (Table \ref{hyadestab}).
These objects are plotted on figure \ref{mg_hyades} as well as the \citet{jameson08, hogan08} and \citet{bannister07} dwarfs.

\begin{table*}
\caption{\label{hyadestab}Name, spectral type, $J$, $H$, $K$ magnitudes, $\Delta\theta$, d$_{mg}$/d$_{sp}$ and d$_{sp}$ for the potential Hyades moving group members discussed.Where two spectral types have been given, they are in the order of optical spectral type, infrared spectral type.}
\begin{center}
\begin{tabular}{l c c c c c c c}
\hline
Name&spt&$J$&$H$&$K$&$\Delta\theta$&d$_{sp}$&d$_{mg}$/d$_{sp}$\\
&&&&&$^{\circ}$&pc&pc\\
\hline
J0131+38&L4& 14.68$\pm$ 0.03& 13.70$\pm$ 0.03& 13.05$\pm$ 0.03&12.18$\pm$1.93 &93.15 & 1.19\\
J0141+18&L1/L4& 13.88$\pm$ 0.03& 13.03 $\pm$0.02& 12.49$\pm$ 0.03&12.65$\pm$2.21& 94.33 & 0.96\\
J0205+12&L5& 15.68$\pm$ 0.06& 14.45$\pm$ 0.05& 13.67$\pm$ 0.04&9.19$\pm$1.79& 91.28 & 0.93\\
J02081+25&L1& 13.99$\pm$ 0.03& 13.11$\pm$ 0.03& 12.59$\pm$ 0.03&12.19$\pm$2.47&93.72 & 1.01\\
J0357-06&L0& 15.95$\pm$ 0.08& 15.06$\pm$ 0.09& 14.60$\pm$ 0.09&5.01$\pm$10.67& 78.52 & 1.02\\
J0624-45&L5& 14.48$\pm$ 0.03& 13.34$\pm$ 0.03& 12.60$\pm$ 0.02& 8.69$\pm$ 5.41&  15.41& 1.27\\
J1928-43&L5& 15.20$\pm$ 0.04& 14.13$\pm$ 0.04& 13.46$\pm$ 0.04& 6.70$\pm$ 3.74 & 21.46& 1.04\\

\hline
\end{tabular}
\end{center}
\end{table*}
\subsection*{Ursa Major}

\begin{figure*}
\begin{center}
\scalebox{0.50}{\includegraphics[angle=270]{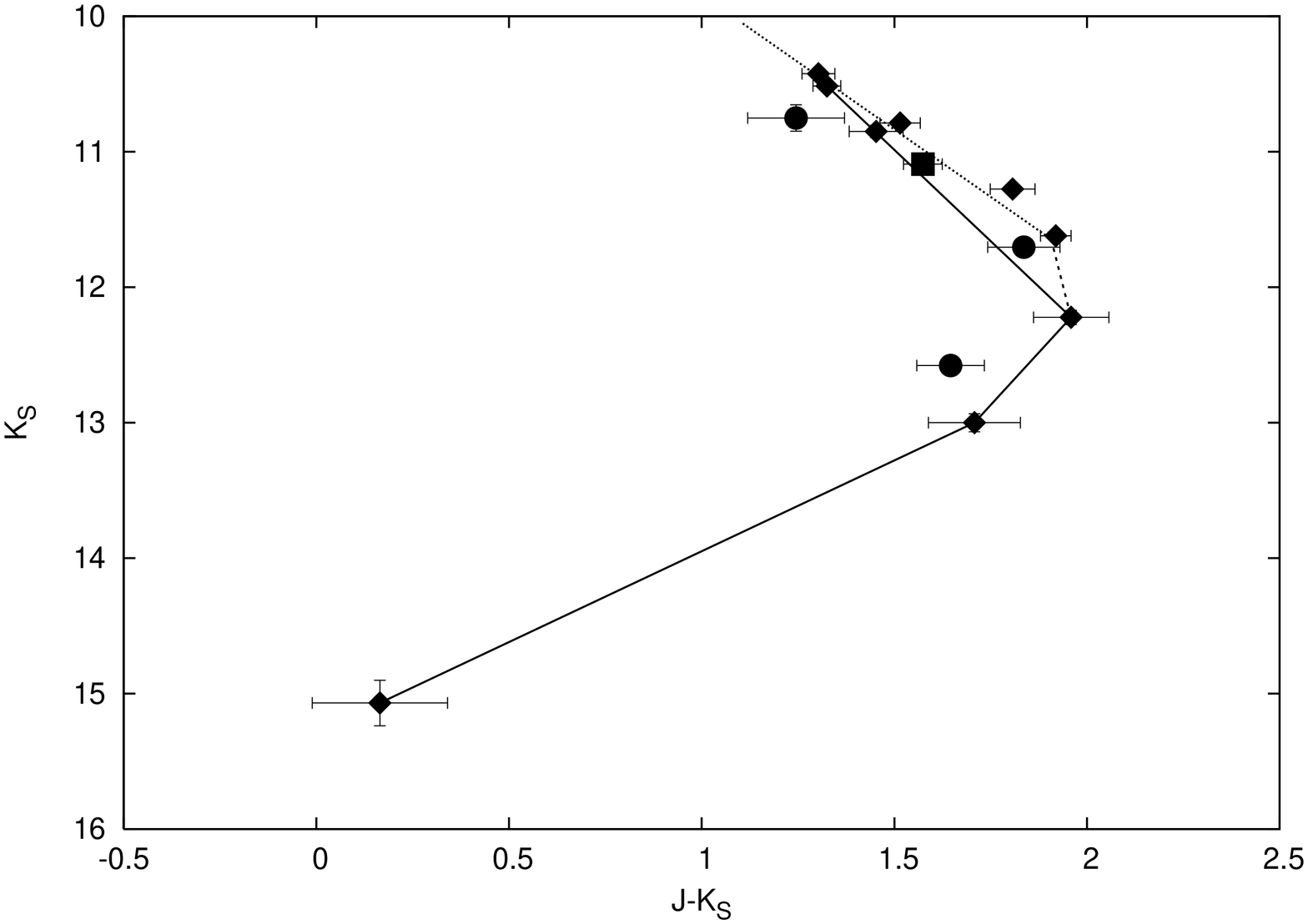}}
\caption{
\label{mg_ursa}
M$_{\rm K}$, $J$-$K$ colour magnitude diagram for the Ursa Major moving group. The cluster members members identified by \citet{jameson08}
 and \citet{bannister07} are marked by filled diamonds.   The selected members are marked by a box for objects from the SofI data and by
 filled circles for the WFCAM objects. The errors are poissonion and from the photometry only. The solid line indicates a possible single 
star sequence from \citet{bannister07}. The straight line fit to the new L dwarf sequence is shown by a dotted line and the dashed line represents what we believe to be the turnaround of the sequence.}
\end{center}
\end{figure*}

The Ursa Major moving group has been estimated to have an age of between 300 Myr \citep{soderblom93} and 500$\pm$100 Myr \citep{king03}.
\citet{castellani02} found an age of the group to be 400 Myr. The age of 400$\pm$100 Myr is adopted in this work.
The convergent point of the Ursa Major moving group is located at $\alpha$=20$^{\rm h}$18$^{\rm m}$.83, $\delta$=$-$34$^{\circ}$25'.8
 (J2000, \citealt{madsen02}).
The selected objects are shown in Figure \ref{mg_ursa} and their data in Table \ref{ursatab}. \\
In \citet{jameson08b} we defined a relationship between M$_{K}$ and $J-K$ for the L dwarf sequence in the Hyades and Ursa Major moving groups and the Pleiades, Upper Scorpius and Alpha Persei open star clusters. This allowed a photometric relationship for estimating the age of any field dwarf to be derived. 

Using the new objects presented here and in Paper I \citep{jameson08} we are able to refine the relationship derived in \citep{jameson08b} for the Ursa Major moving group.
Previously, this relationship  (which takes the form $J-K$=m $\times M_K$+c, in the MKO system) had a gradient (m)=2.88 and c=11.22. This was only based on the 2 brown dwarfs in the L dwarf sequence from \citet{bannister07}, and was obviously incorrect. The sequence shown in Figure 2 of \citet{jameson08b} crossed that for the Hyades, and the gradient of 2.88, did not fit, with the gradients of the other clusters, which had approximately parallel sequences, all with gradients of $\sim$1.98, the value that was suggested by the authors as being more appropriate.
Using our new L dwarfs, and those found by \citet{jameson08}, we used the colour conversions of \citet{stephens04} to recalculate the relationship on the MKO system.
Our new values are m=2.0746$\pm$0.281, c=7.92052$\pm$0.4481, which is more in keeping with the values for the other clusters.
The relationship in the 2MASS colour system is plotted on Figure \ref{mg_ursa}, as well as the sequence as defined by the \citet{bannister07} dwarfs.

\begin{table*}
\caption{\label{ursatab}Name, spectral type, $J$, $H$, $K$ magnitudes, $\Delta\theta$, d$_{mg}$/d$_{sp}$ and d$_{sp}$ for the potential
 Ursa Major moving group members discussed.Where two spectral types have been given, they are in the order of optical spectral type, 
infrared spectral type.J0920+35 is a binary \citep{reid01}, discovered using $HST$. Is suspected to be an equal mass binary.}
\begin{center}
\begin{tabular}{l c c c c c c c}
\hline
Name&spt&$J$&$H$&$K$&$\Delta\theta$&d$_{sp}$&$_{mg}$/d$_{sp}$\\
&&&&&$^{\circ}$&pc\\
\hline
J0218-31&L3& 14.73$\pm$0.04& 13.81$\pm$0.04& 13.15$\pm$0.03& 6.29$\pm$ 5.14 &25.85& 0.86\\
J0310-27&L5& 15.80$\pm$ 0.07 &14.66$\pm$ 0.05& 13.96$\pm$ 0.06& 1.48 $\pm$7.36&  28.24& 0.80\\
J0815+45 &L1&16.06$\pm$ 0.08& 15.23$\pm$ 0.09& 14.81$\pm$ 0.10& 12.04$\pm$ 15.89&  64.88& 1.09\\
J0920+35&L6.5/T0& 15.63$\pm$ 0.06& 14.67$\pm$ 0.06& 13.98$\pm$ 0.06& 5.30$\pm$ 2.96&  19.06& 0.79\\
\hline
\end{tabular}
\end{center}
\end{table*}
\subsection*{Pleiades}
\begin{figure*}
\begin{center}
\scalebox{0.50}{\includegraphics[angle=270]{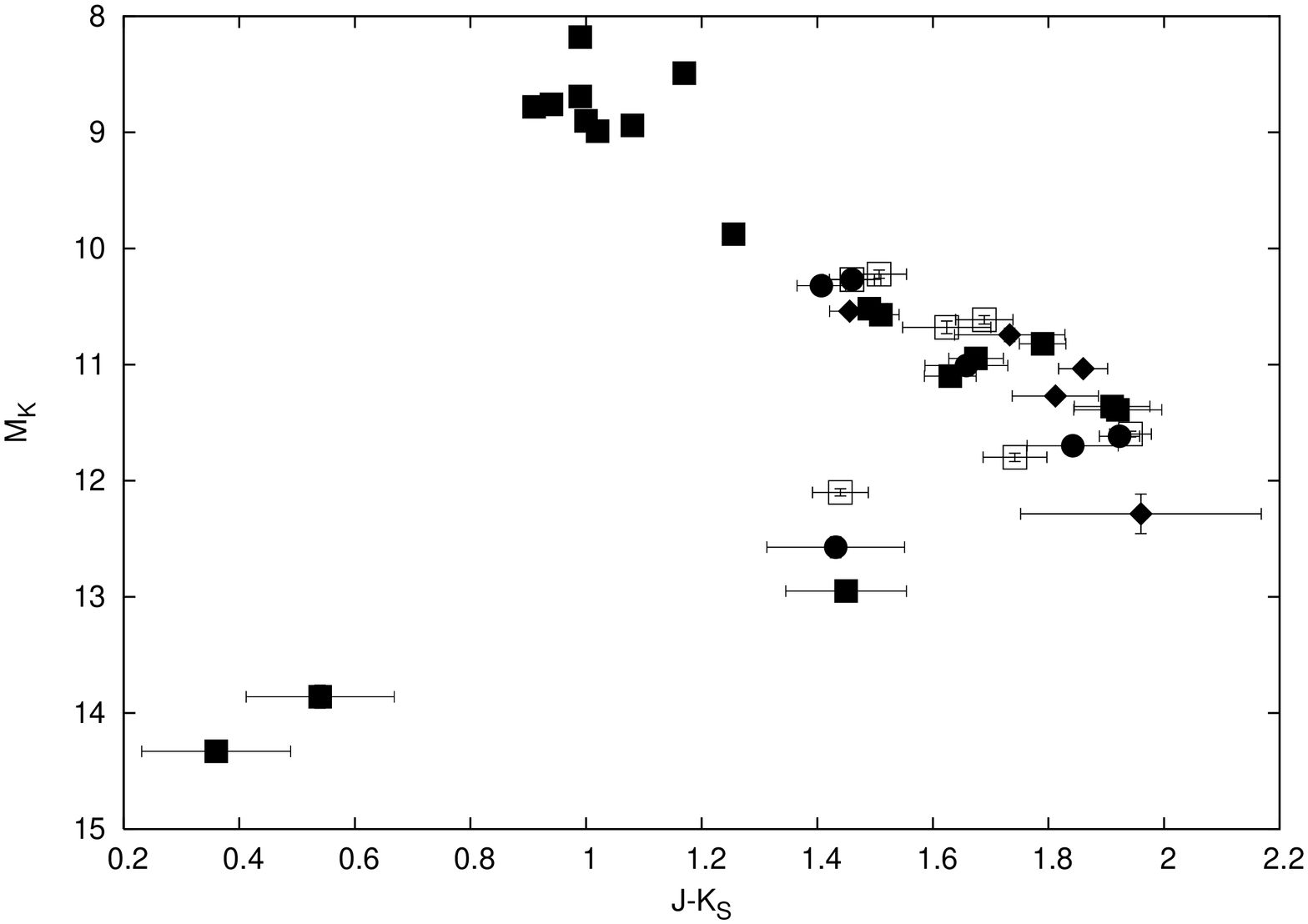}}
\caption{
\label{mg_plds}
M$_{\rm K}$, $J$-$K$ colour magnitude diagram for the Pleiades moving group. The cluster members members identified by \citet{casewell07}, \citet{lodieu07a}, \citet{moraux03} and \citet{bihain06} are plotted as filled squares.  All of the objects that were selected as moving group members are marked as filled diamonds \citep{jameson08}. The selected members are marked by a box for objects from the SofI data and by filled circles for the WFCAM objects. The errors are poissonion and from the photometry only.}
\end{center}
\end{figure*}

The Pleiades cluster is 125 Myr old and is situated at a distance of 130 pc \citep{stauffer98}.
As a cluster it has been studied in depth and has been found to contain many brown dwarfs \citep{casewell07, lodieu07a,bihain06,moraux03}. The Pleiades moving group has a convergent point of 85.04$^{\circ}$$\pm$3.67, $-$39.11$^{\circ}$$\pm$6.92 \citep{madsen02}. This convergent point is very close to that of many other moving groups such as  Alpha Persei (96$^{\circ}$.78$\pm$1.96, -23$^{\circ}$.27$\pm$3.67; \citealt{madsen02}, 50 Myr; \citealt{lynga87}), Tucana/Horologium (30 Myr; \citealt{zuckerman04}) and the AB Dor moving group (50 Myr; \citealt{zuckerman04}) (see \citealt{zuckerman04} for a review), and it has been theorised that many of these moving groups have a common origin \citep{ortega07}.
Thirteen new candidate members were found using the moving group method. 6 objects with declinations $<$-30$^{\circ}$ and 7 objects with declinations $>$-30$^{\circ}$ (Figure \ref{mg_plds}, Table \ref{pleiadestab}).
As with the Hyades and Ursa Major clusters, the Pleiades is old enough to expect that some mass segregation has occurred, and thus it is not unreasonable to search the whole sky for members of the moving group. This is not the case however for many of the younger, southern moving groups.

The new moving group members are detailed in Table \ref{pleiadestab}.

\begin{table*}
\caption{\label{pleiadestab}Name, spectral type, $J$, $H$, $K$ magnitudes, $\Delta\theta$, d$_{mg}$/d$_{sp}$ and d$_{sp}$ for the potential Pleiades moving group members discussed.Where two spectral types have been given, they are in the order of optical spectral type, infrared spectral type.}
\begin{center}
\begin{tabular}{l c c c c c c c}
\hline
Name&spt&$J$&$H$&$K$&$\Delta\theta$&d$_{sp}$&d$_{mg}$/d$_{sp}$\\
&&&&&$^{\circ}$&pc\\
\hline
J0032-44&L0& 14.78$\pm$ 0.03& 13.86$\pm$ 0.03& 13.27$\pm$ 0.04& 13.71$\pm$ 2.51&  40.70& 0.86\\
J0058-06&L0& 14.31$\pm$ 0.03& 13.44$\pm$ 0.03& 12.90 $\pm$0.03& 4.83$\pm$ 5.19 & 32.85 & 0.99\\
J0117-34&L2& 15.18$\pm$ 0.03& 14.20$\pm$ 0.04& 13.49$\pm$ 0.04& 4.84$\pm$ 4.28&  37.58& 1.18\\
J0125-34&L2& 15.52$\pm$ 0.05& 14.47$\pm$ 0.05& 13.90$\pm$ 0.05& 18.82$\pm$ 10.14&  44.03& 0.90\\
J0144-07&L5& 14.20$\pm$ 0.03& 13.01$\pm$ 0.03& 12.27$\pm$ 0.02& 13.87$\pm$ 2.28 &  13.49 & 1.02\\
J0208+27&L5& 15.71$\pm$ 0.06& 14.56$\pm$ 0.06& 13.87$\pm$ 0.05& 15.56$\pm$ 3.60 &  27.21  &0.903\\
J0236+00&L6/L6.5& 16.10$\pm$ 0.08& 15.27$\pm$ 0.07& 14.67 $\pm$0.09& 8.13$\pm$ 3.58 & 26.22 & 1.00\\
J0316-28&L0& 14.57$\pm$ 0.04& 13.77$\pm$ 0.03 &13.11$\pm$ 0.03& 4.55$\pm$ 7.30 &  37.02 & 1.17\\
J0357-44&L0& 14.37 $\pm$0.03& 13.53$\pm$ 0.03& 12.91$\pm$ 0.03& 4.12$\pm$ 6.85&  33.71& 0.99\\
J0409+21&L3& 15.51$\pm$ 0.05& 14.50$\pm$ 0.05& 13.85$\pm$ 0.05& 16.18$\pm$ 4.55 &  37.03 & 0.94\\
J1425-36&L5& 13.74 $\pm$0.03& 12.57$\pm$ 0.02& 11.81$\pm$ 0.03& 6.23$\pm$ 2.76&  11.00& 1.11\\
J1928-43&L5& 15.20 $\pm$0.04& 14.13$\pm$ 0.04& 13.46$\pm$ 0.04& 3.80$\pm$ 3.74&  21.46& 1.05\\
J1936-55&L5& 14.49 $\pm$0.04& 13.63$\pm$ 0.03& 13.05$\pm$ 0.03& 15.30$\pm$ 4.84&  15.46& 1.17\\

\hline
\end{tabular}
\end{center}
\end{table*}
\subsection*{Other moving groups}
Many of the southern moving groups have similar convergent points and velocities. To determine if any of our southern dwarfs are members of these moving groups, we have used the moving group method as for the Hyades, Ursa Major 
and Pleiades moving groups, but have then used the isochrone for Upper Scorpius as developed by \citet{jameson08b}. This can be then used as an age indicator
for the younger clusters. If the selected dwarfs fall on or near the isochrone, then they are young (age $<$10 Myr) and are considered candidate members.
Radial velocity measurements are needed to confirm the membership of these objects however.

Simply by using the moving group method, we have found $<$10 candidate members of the TW Hydra, Tucana/Horlogium, Beta Pictoris, AB Doradus and $\eta$ Chamaeleon
moving groups. Because these clusters all have similar convergent points and velocities \citep{zuckerman04} some of the candidate members are found in more than
one cluster.

When plotted on the M$_K$, $J-K$ colour magnitude diagram, using \citet{stephens04} to convert the colours into the MKO system, with 
the Upper Scorpius and Alpha Perseus cluster isochrones from \citet{jameson08b}, all of the candidate members sit lower then Alpha Per, indicating that 
these dwarfs cannot be moving group members at the calculated distances.

\section*{Conclusions}
This paper continues the work presented in \citet{jameson08}, which presented proper motions of 143 L and T dwarfs. This paper presents a further 126 proper motions.
Thus the large majority of field L and T dwarfs discovered by 2MASS, DENIS and the SDSS now have known proper motions.
From these data we find a further 2 wide binary L dwarfs, both with M dwarf companions.
A further 3 high velocity L dwarfs have been discovered, which we assume are old thick disc L dwarfs.  Finally, we find 7 more potential members of the Hyades moving group, 4 members of the Ursa Major moving group and 13 potential members of the Pleiades moving group. We have found no members of the young southern moving groups, TW Hydra, Tucana/Horologium, Beta Pictoris, AB Doradus or $\eta$ Chamaeleon. 
We have also used these new members of the Ursa Major moving group to refine the L dwarf sequence for the group that was defined by \citet{jameson08b}.

\section*{Acknowledgements}
SLC was supported by STFC for the duration of this work. MBU is supported by a STFC Advanced Fellowship.
Observations were made  at the
 United Kingdom Infrared Telescope, which is operated by the Joint Astronomy Centre on behalf of the U.K. Particle Physics and Astronomy Research Council.Observations were also made a the New Technology Telescope which is operated by ESO.
This publication makes use of data products from the Two Micron All Sky Survey, which is a joint project of the University of Massachusetts and the Infrared Processing and Analysis Center/California Institute of Technology, funded by the National Aeronautics and Space Administration and the National Science Foundation.
Research has benefited from the M, L, and T dwarf compendium housed at DwarfArchives.org and maintained by Chris Gelino, Davy Kirkpatrick, and Adam Burgasser.
This research has made use of NASA's Astrophysics Data System Bibliographic Services.

\label{lastpage}

\end{document}